\documentclass[%
 reprint,
superscriptaddress,
 amsmath,amssymb,
 aps,
]{revtex4-2}
\usepackage{soul}
\usepackage{graphicx}
\usepackage{dcolumn}
\usepackage{bm}
\usepackage{braket}
\usepackage{amsmath}
\usepackage{bbold}
\usepackage{xcolor}
\usepackage{amsthm}
\usepackage{physics}
\usepackage{enumitem}
\usepackage{centernot}

\usepackage{hyperref}
\hypersetup{colorlinks=true,linkcolor=teal,citecolor=teal,urlcolor=teal}

\theoremstyle{remark}

\begin{document}

\preprint{APS/123-QED}


\title{Catalytic advantage in Otto-like two-stroke quantum engines}

\author{Marcin {\L}obejko}
\affiliation{Institute for Theoretical Physics and Astrophysics, Faculty of Mathematics,
Physics and Informatics, University of Gda\'nsk, 80-308 Gda\'nsk, Poland}
\affiliation{International Centre for Theory of Quantum Technologies, University of Gdansk, Wita Stwosza 63, 80-308 Gdansk, Poland}

\author{Tanmoy Biswas}
\affiliation{International Centre for Theory of Quantum Technologies, University of Gdansk, Wita Stwosza 63, 80-308 Gdansk, Poland}
\affiliation{Theoretical Division (T4), Los Alamos National Laboratory, Los Alamos, New Mexico 87545, USA.}

\author{Pawe{\l} Mazurek}
\affiliation{Institute of Informatics, Faculty of Mathematics, Physics and Informatics, University of Gdansk, Wita Stwosza 63, 80-308 Gdansk, Poland}
\affiliation{International Centre for Theory of Quantum Technologies, University of Gdansk, Wita Stwosza 63, 80-308 Gdansk, Poland}

\author{Micha{\l} Horodecki}
\affiliation{International Centre for Theory of Quantum Technologies, University of Gdansk, Wita Stwosza 63, 80-308 Gdansk, Poland}

\date{\today}

\begin{abstract}
We demonstrate how to incorporate a catalyst to enhance the performance of a heat engine. Specifically, we analyze efficiency in one of the simplest engines models, which operates in only two strokes and comprises of a pair of two-level systems, potentially assisted by a $d$-dimensional catalyst. When no catalysis is present, the efficiency of the machine is given by the Otto efficiency. Introducing the catalyst allows for constructing a protocol which overcomes this bound, while new efficiency can be expressed in a simple form as a generalization of Otto's formula: $1 - \frac{1}{d} \frac{\omega_c}{\omega_h}$. The catalyst also provides a bigger operational range of parameters in which the machine works as an engine. Although an increase in engine efficiency is mostly accompanied by a decrease in work production (approaching zero as the system approaches Carnot efficiency), it can lead to a more favorable trade-off between work and efficiency. The provided example introduces new possibilities for enhancing performance of thermal machines through finite-dimensional ancillary systems.


\end{abstract}

\maketitle

 With inspiration coming from the field of chemistry, a catalyst has been introduced in quantum information theory as an auxiliary system that expands the range of possible state transformations while remaining unchanged throughout the protocol \cite{JonathanPlenio99,vanDamHayden2007}. Catalysis was subsequently also applied to quantum thermodynamics \cite{brandao2015second,Aberg2014,Ng2015,MullerPastena2016,LostaglioMueller2015,Muller2018PRX,Boes2019PRL,ShiraishiSagawa21,WilmingPRL,Wilming2022correlationsin,CDatta_Review,Bartosik_review}. 
 In particular, catalysis was used to enhance performance of cooling  \cite{Henao2021catalytic,HenaoUzdin}.
In the context of thermal machines working in a cyclic fashion, it emerges as the most apparent generalized thermodynamic resource.

In this Letter, we want to characterize the catalyst's ability to enhance the
performance of quantum heat engines. We show that the presence of a catalyst not only boosts the efficiency, but also 
can drive a machine that was not working as an engine to the engine regime. Moreover, the catalyst can also boost engine's power (i.e., work per cycle) or lead to a more favorable power vs. efficiency trade-off.   These results are obtained within a class of two-stroke engines closely related to the Otto engine. 

The conventional Otto cycle is one of the most well-known paradigms for constructing heat engines at the microscopic scale and has been thoroughly investigated both theoretically \cite{Scovil1959,Scovil2,Alicki1979,Kosloff1984,Kosloff_Rezek,Henrich2007,Feldmann_Kosloff,Feldmann_Kosloff_finite,Discrete_Kosloff,Nori,Zhang_Meystre,Wang,Myers2022,PopescuSmallestHeatEngine2010,PopescuSmallestHeatEngineprinciple2010,Hofer_2017} and experimentally \cite{Rossnagel2016,Blickle2012,Klaers_nanobeam,de_Assis,Peterson_Serra,Maslennikov2019,Myers2022}. Its simple operational mode is based on interaction with the environment alternating with energy level transformations, 
collectively forming the four-stroke thermodynamic cycle. The central question in the Otto engine studies is how to enhance its performance. Most of the research primarily focuses on dynamical optimization, aiming to achieve a better balance between power and efficiency by considering different system-environment couplings \cite{Beau_delcampo,Chen,de_Assis,Camati_Serra}. Yet, there has been limited attention given to fundamentally improving efficiency alone, solely through a less dissipative heat-to-work conversion. In this case, there seems to be no novel ideas explored beyond the obvious approaches like the usage of the non-thermal baths \cite{Niedenzu_2016,Niedenzu2018,Thomas_Ghosh,XIAO20183051,Manzano}.

We consider an operational simplification of the Otto cycle, which reduces it from four to just two strokes \cite{two-stroke-allahverdyan,ClivazPRL,ClivazPRE,Landauer_vs_Nernst,Lostaglio,Otavio,BrunnerVirtualqubit2012,Silvadimension}.
The fundamental concept of the two-stroke engine involves breaking down the cycle into two distinct stages: (i) work extraction through an isoentropic process and (ii) heat exchange via thermalization with heat baths. This division significantly streamlines the engine's operation, as after the work-stroke, the cycle can be promptly completed by bringing the working body into contact with the respective environments. 
The primary trade-off for this simplification is an increase in the engine's dimension. Specifically, the  simplified, two-stroke engine requires simultaneous operation on the pair of two-level systems (TLSs), while only a single TLS is required for the four-stroke Otto engine. 
Here, we focus on the two stroke-model representing an analogue of the Otto cycle.
However, the concept of two-stroke engines extends beyond this, and detailed analysis, including optimality proofs, will be presented for this general model in the accompanying paper \cite{BiswasNew}, 

Within this paradigm, we provide a systematic methodology how to include a $d$-dimensional catalyst into an operation of the two-stroke engine that leads to the generalized ``$d$ - Otto'' efficiency, given by the formula: 
\begin{eqnarray}
    \eta_d = 1 - \frac{1}{d} \frac{\omega_c}{\omega_h},
\end{eqnarray}
with a positive work production for 
\begin{align}
d\in \left(\frac{\omega_c}{\omega_h},\frac{\beta_c\omega_c }{\beta_h\omega_h}\right).
\end{align}
We identify regions of parameters (temperatures $\beta_c$
and $\beta_h$ and frequencies $\omega_c$ and $\omega_h$) where presence of the catalyst is a neccessary condition for the machine to work as an engine. In other regimes, he catalyst brings higher efficiency and work production.

Although we concentrate here on engines, the idea can be easily generalized and applied for refrigerators and heat pumps as well. Generally, it opens a new range of possibilities for increasing the performance of thermal machines. Notably, in contrast to most of the other studies, in our case, the catalyst is included explicitly as the finite-dimensional system.

\begin{figure}
    \centering
    \includegraphics[width=0.45\textwidth]{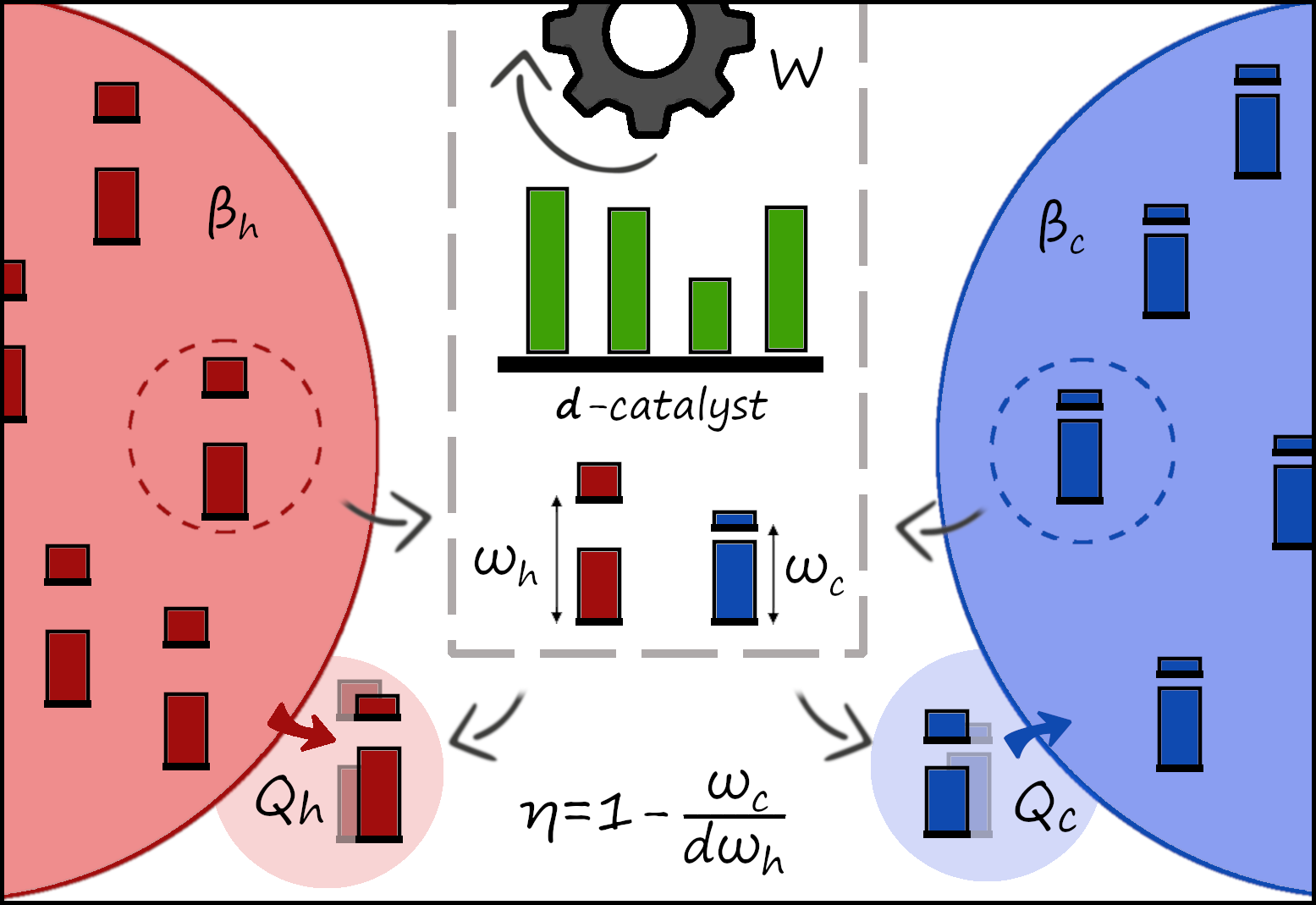}
    \caption{Two-stroke engine assisted by a catalyst. The operational principle of a two-stroke engine is to combine thermal resources, such as hot and cold thermal two-level systems, and use an assisted catalyst to reduce the total energy of the system by performing work $W$ on an external agent. The first work-stroke is performed on an isolated system and is considered a unitary process. After that, the used fuels are removed and thermalized again, triggering corresponding hot $Q_h$ and cold $Q_c$ heat flows. The catalyst, after removing the fuel, returns to its initial state, allowing the entire process to be repeated. As a main result, we demonstrate that a $d$-dimensional catalyst can enhance the efficiency of the process to the generalized Otto's formula: $\eta = 1 - \frac{1}{d} \frac{\omega_c}{\omega_h}.$}
    \label{fig:self-contained engine}
\end{figure}

\emph{Two-stroke engine.} We commence by introducing a two-stroke engine, which is conceptually depicted in Figure \ref{fig:self-contained engine}. The engine operates in two fundamental steps: the work-stroke and the heat-stroke. In the initial stage, two thermal systems denoted as $\tau_h$ and $\tau_c$ (coming from hot and cold bath, respectively) are brought into contact with a third system labeled as $\rho_s$, functioning as a catalyst. The work-stroke, in general, involves the implementation of the cyclic isoentropic process, which serves the purpose of reducing the overall energy of the composite system. Here, we concentrate solely on the unitary processes, such that together with cylicity condition, the work-stroke corresponds to the ergotropy extraction process \cite{Allahverdyan_2004,Lobejko2020,Lobejko2021,BiswasQuantum,Tirone_PRL}. From a physical standpoint, due to unitarity and energy conservation, we interpret this process as work extraction carried out by an external agent. In the final stage, i.e., the heat-stroke, the hot and cold systems are once again thermally equilibrated, which also uncorrelates the catalyst and restores it to its initial state. With this construction the thermodynamic cycle is completed, and the engine is ready to repeat the whole process again.

A mathematical description of the proposed engine is based on the following assumptions: (i) initial state of the engine is defined as the product state: $\varrho = \tau_{h} \otimes \tau_c \otimes \rho_s$, where $\tau_{h,c} \propto \exp(-\beta_{h,c} H_{h,c})$ are thermal states with corresponding (inverse) temperatures $\beta_{h} < \beta_c$ and Hamiltonians $H_{h,c}$; (ii) a single cycle of the engine is described by a unitary process $U$, such that $\varrho \to U \varrho U^\dag$; (iii) the final marginal state of the catalyst is equal to the initial one:
    $\Tr_{hc}[ U \varrho U^\dag] = \rho_s.$
Then, thermodynamics of the engine is introduced based on the following exchanged heat definitions:
\begin{eqnarray}
    Q_{h,c} &=& \Tr[H_{h,c} (\varrho-U \varrho U^\dag)], 
\end{eqnarray}
which is precisely the amount of energy that needs to be provided by external baths in order to thermalize the systems. In accordance with the First Law, the work that is provided by external agent is equal to $W = Q_c + Q_h$, such that efficiency of the engine is given by:
\begin{eqnarray}
    \eta = \frac{W}{Q_h} = 1 + \frac{Q_c}{Q_h}.
\end{eqnarray}
One may prove that in this framework the Second Law is satisfied, such that the Clausius inequality holds: $\beta_h Q_h + \beta_c Q_c \le 0$. 

In this Letter, we are interested in one of the simplest self-contained engines that is consisted of the hot $\tau_h$ and cold $\tau_c$ two-level systems (TLSs), and the $d$-dimensional catalyst $\rho_s$. The initial thermal states of TLSs are $\tau_k = \mathcal{Z}_k^{-1} (\dyad{0}_k + e^{-\beta_k \omega_k} \dyad{1}_k)$ with Hamiltonians $H_k = \omega_k \dyad{1}_k$, where $\mathcal{Z}_k = 1+e^{-\beta_k \omega_k}$ (for $k = h,c$). The catalyst is described by the density matrix $\rho_s = \sum_i p_i \dyad{i}_s$ with an arbitrary Hamiltonian. Moreover, we concentrate solely on unitary operations $U$ given by the set of transpositions of the energy levels (the so-called swaps). For example: the swap $\ket{ijk} \leftrightarrow \ket{i'j'k'}$, corresponds to the unitary action $U\ket{ijk} = \ket{i'j'k'}$ and $U \ket{i'j'k'} = \ket{ijk}$.

\begin{figure*}
    \centering
    \includegraphics[height = 0.5\textwidth]{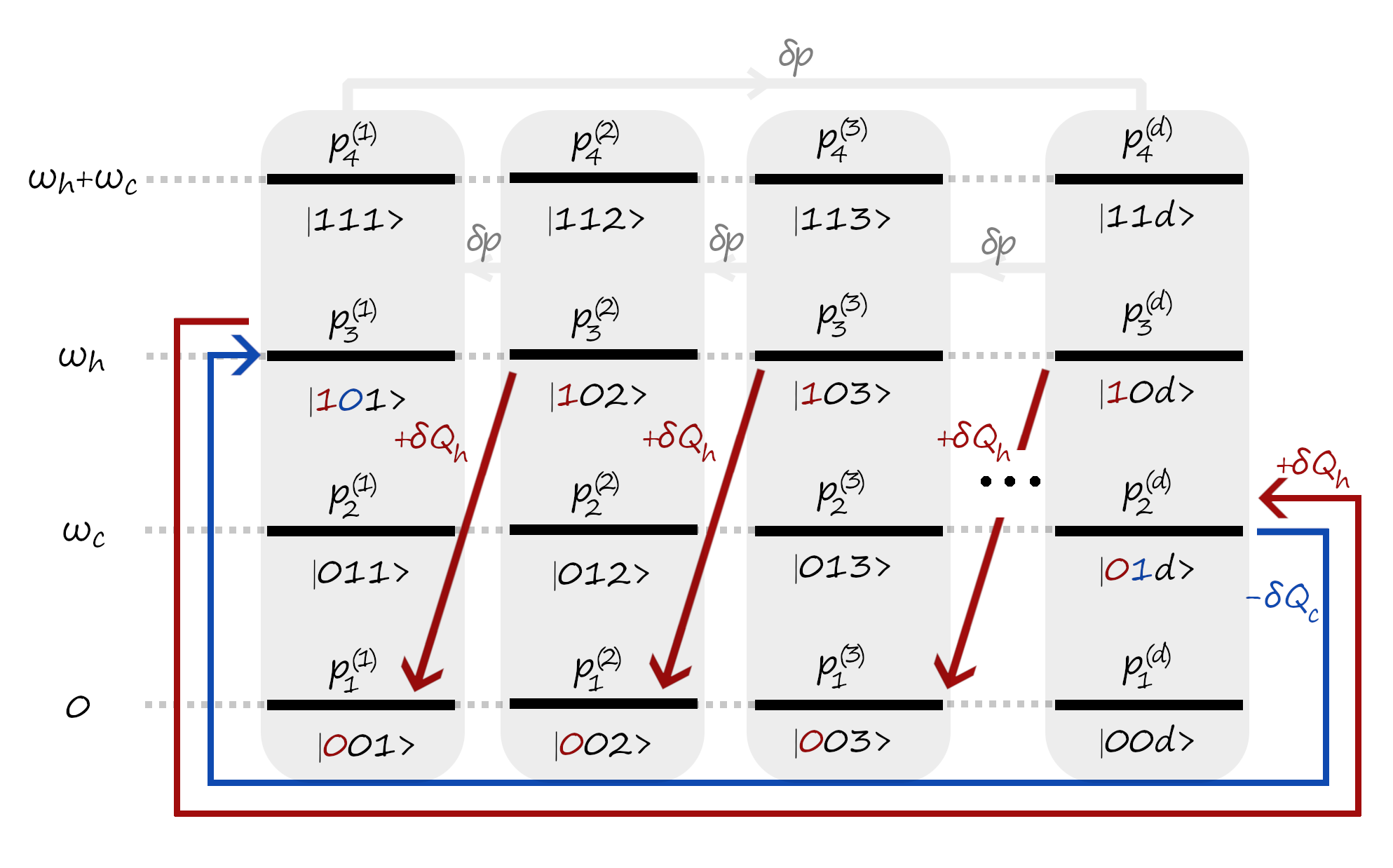}
    \caption{Graphical representation of disjoint swaps that lead to $d$ -  Otto efficiency. The plot represents all the energy levels of the composite system, where $\ket{i j k } \equiv \ket{i}_h \ket{j}_c \ket{k}_s$. Each column represent a $k$-th state of the catalyst, such that cyclicity condition reflects the conservation of the sum of probabilities within the column. For disjoint swaps this boils down to the equal flow of the probability $\delta p$ between the connected columns.  Connected levels (either via red or blue arrow) are swapped within the work extraction stage. The red and blue arrows represents the energy changes of hot and cold TLS, respectively, where the up-arrow correspond to increase and down-arrow to decrease in energy. The process is designed in such a way that if $\delta p > 0$, the hot TLS is deexcited in each transposition 
    and the cold TLS is excited in one of the transposition.
    For $d$ transpositions, 
    we get the following heat flows: $Q_h = d \delta Q_h =  d \omega_h \delta p$ and $Q_c = - \delta Q_c = - \omega_c \delta p$, which implies $\eta = 1 + Q_c/Q_h = 1 - \omega_c /(d \omega_h)$. }
    \label{fig:d_swaps}
\end{figure*}

\emph{Otto efficiency (without a catalyst).} Let us start with a protocol with two TLSs (without the assisted catalyst), which leads to the Otto efficiency. In this case, to extract positive work $W > 0$ via the transposition of the energy levels, there must be an inversion of the population in the composite state of $\tau_h \otimes \tau_c$. In the following, we represent the initial (diagonal) density matrix by the vector of probabilities $\vec p_0 = (\mathcal{Z}_h \mathcal{Z}_c)^{-1} (1, e^{-\beta_h \omega_h}, e^{-\beta_c \omega_c}, e^{-\beta_h \omega_h -\beta_c \omega_c})$ (with corresponding energies $0, \omega_h, \omega_c$ and $\omega_h + \omega_c$).
From here we see that the ground state $\ket{00}$ is the most occupied state and the excited state $\ket{11}$ is the least. Thus, the only possibility to extract work from the system via swap is to transpose $\ket{01} \leftrightarrow \ket{10}$ providing that $e^{-\beta_h \omega_h} > e^{-\beta_c \omega_c}$ and $\omega_h > \omega_c$ (i.e., the inversion of population is present). In this case, we achieve the Otto efficiency:
\begin{eqnarray}
    \eta_1 = 1 - \frac{\omega_c}{\omega_h}
\end{eqnarray}
with the work production:
\begin{eqnarray}
    W_1 = \frac{(\omega_h - \omega_c)(e^{-\beta_h \omega_h} - e^{-\beta_c \omega_c})}{(1+e^{-\beta_h \omega_h})(1+e^{-\beta_c \omega_c})}.
\end{eqnarray}
One may also prove that this is the optimal efficiency in the whole set of unitary operations \cite{BiswasNew}.


\emph{2 - Otto efficiency (two-dimensional catalyst).} Let us now reveal how the efficiency of the process may be increased via the assisted catalyst, which essentially boils down to increasing the hot heat $Q_h$ while keeping the cold one $Q_c$ constant.

From now on, we concentrate solely on disjoint swaps. Then, one may consider an \emph{internal} swap $\ket{ijk} \leftrightarrow \ket{i'j'k}$, that does not affect the state of the catalyst, or the \emph{external} swap $\ket{ijk} \leftrightarrow \ket{i'j'k'}$, with the change in the catalyst state. Clearly, the protocol based solely on internal swaps satisfies the cyclicity at the prize of trivializing the problem. Thus, we consider at least one external swap of the type $\ket{ijk} \leftrightarrow \ket{i'j'k'}$. Now, let us label the initial set of occupation probabilities by $\vec p_l = (p_1^{(l)}, p_2^{(l)}, p_3^{(l)}, p_4^{(l)})$, for $l=1,2$ representing two different states of the catalyst, such that $\vec p_1 = p \ \vec p_0$ and $\vec p_2 = (1-p) \ \vec p_0$, where $p \in [0,1]$ describes the state of the catalyst. The final state of the system (after the permutation) is labeled by $\vec{s}_l = (s_1^{(l)}, s_2^{(l)}, s_3^{(l)}, s_4^{(l)})$, such that the cyclicity constraint is given by $\sum_i p_i^{(l)} = \sum_i s_i^{(l)}$.

Then, let us consider a protocol with only one external swap. We label it by $i \leftrightarrow j$, when the $i$th state from the first block is swapped with the $j$th state from the second one. E.g. a swap $1 \leftrightarrow 3$ results with a set of occupation probabilities: $\vec{s}_1 = (p_3^{(2)}, p_2^{(1)}, p_3^{(1)}, p_4^{(1)})$ and $\vec{s}_2 = (p_1^{(2)}, p_2^{(2)}, p_1^{(1)}, p_4^{(2)})$. In general, for the $i \leftrightarrow j$ swap, one can show that the heat flow is given by: 
\begin{eqnarray}
    Q_{h,c} = (\varepsilon_i^{h,c} - \varepsilon_j^{h,c}) (p_i^{(1)} - p_j^{(2)}),
\end{eqnarray}
where $\varepsilon_i^{h,c}$ are the corresponding energies of the hot or cold TLS associated with the $i$th state. However, one can easily show that to satisfy the cyclicity condition, the equality $p_i^{(1)} = p_j^{(2)}$ has to be obeyed, which consequently results in no heat and work flow at all. The same reasoning is true for the process with three external swaps since the blocks for $k=1$ and $k=2$ are equivalent, and the three-swap process is equivalent to the one-swap process. 

As a conclusion, we see that the only advantage that we can get from the presence of the catalyst is for a process with two external swaps. Let us then consider a two disjoint swaps: $i \leftrightarrow j$ and $n \leftrightarrow m$. From this we get a formula for the heat flow:
\begin{eqnarray}
    Q_{k} = (\varepsilon_i^k - \varepsilon_j^k) (p_i^{(1)} - p_j^{(2)}) + (\varepsilon_n^k - \varepsilon_m^k) (p_n^{(1)} - p_m^{(2)}),
\end{eqnarray}
and the cyclicity condition translates to the relation:
\begin{eqnarray} \label{cyclic_condition_two_dim}
    p_i^{(1)} + p_n^{(1)} = p_j^{(2)} + p_m^{(2)},
\end{eqnarray}
such that $p_i^{(1)} - p_j^{(2)} = -(p_n^{(1)} - p_m^{(2)}) \equiv \delta p$. Finally, we have
\begin{eqnarray}
    Q_k = (\varepsilon_i^k + \varepsilon_m^k - \varepsilon_j^k - \varepsilon_n^k) \delta p.
\end{eqnarray}
Then, to maximize the efficiency one has to increase the hot heat flow and decrease the cold one. This can be achieved by putting $i= m = 3$ and $j = 2, n = 1$. 
In accordance, we get
\begin{eqnarray}
    Q_h = 2 \omega_h \delta p, \quad Q_c = -\omega_c \delta p,
\end{eqnarray}
which results in the so-called ``$2$ - Otto'' efficiency:
\begin{eqnarray}
    \eta_2 = 1 - \frac{\omega_c}{2\omega_h}.
\end{eqnarray}
To calculate the work production $W = Q_h + Q_c$, one needs to solve the equation \eqref{cyclic_condition_two_dim} for $\delta p$, which results in:
\begin{eqnarray}
    W_2 = \frac{(2\omega_h - \omega_c)(e^{-2\beta_h \omega_h} - e^{-\beta_c \omega_c})}{(1+e^{-\beta_h \omega_h})(1+e^{-\beta_c \omega_c}) f_2},
\end{eqnarray}
where $f_2 = 1 + e^{-\beta_c \omega_c} + 2 e^{-\beta_h \omega_h}$. \\

\begin{figure}
    \centering
    \includegraphics[width=0.90\linewidth]{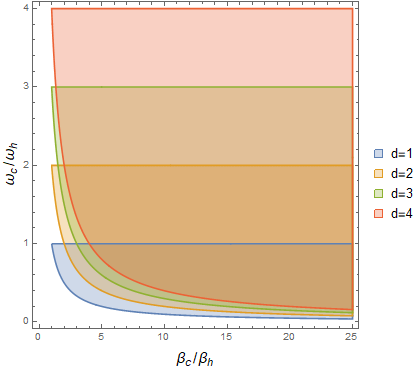}
    \caption{Regimes of the engine mode. According to inequality \eqref{regime_ineq} the regime of the engine mode (i.e., with $W>0$) is provided solely by the frequency ratio $\omega_c/\omega_h$ and temperature ratio $\beta_c/\beta_h$. The graph presents regimes for different dimensions of the catalyst $d$.}
    \label{fig:ranges}
\end{figure}

\emph{$d$ -  Otto efficiency ($d$-dimensional catalyst).} The generalization to $d$-dimensional catalyst with disjoint transpositions is straightforward. To simplify the notation, we parameterize swaps in the following way.   An $i$th swap $u_i \leftrightarrow d_i$ is a transposition of the energy levels $\ket{u_i}$ (``up'') and $\ket{d_i}$ (``down''), such that the corresponding energies satisfy: $\varepsilon_{u_i} > \varepsilon_{d_i}$. In accordance, the heat flows can be written as:
\begin{eqnarray} \label{heat_flow}
    Q_{k} = \sum_i \omega_i^k (p_{u_i} - p_{d_i}),
\end{eqnarray}
where $\omega_i^k = \bra{u_i} H_k \ket{u_i} - \bra{d_i} H_k \ket{d_i}$ and $p_{u_i} = \bra{u_i} \varrho \ket{u_i}$, $p_{d_i} = \bra{d_i} \varrho \ket{d_i}$, while the cyclicity condition is given by:
\begin{eqnarray}
 \sum_i (\delta_{l,u_i} - \delta_{l,d_i}) (p_{u_i} - p_{d_i}) = 0
\end{eqnarray}
for each $l = 1, 2, \dots, d$. Let us note here that within this generalized description, the trivial case $d=1$ refers to no catalyst present. In analogy to the two-dimensional case, the strategy is to
design the swaps in such a way that:  $\omega_i^h = \omega_h$ (for all $i$), $\omega_j^c = -\omega_c$ (for some $j$) and $\omega_i^c = 0$ (for all $i \neq j$). According to this idea, one can easily generalize a two-swap case to the $d$-swap protocol putting:
\begin{align}
    &\ket{u_i} = \ket{10i} \quad \text{for} \quad i=1,2,\dots, d \\
    &\ket{d_i} = \ket{00i} \quad \text{for} \quad i=1,2,\dots, d-1  \\
    &\ket{d_d} = \ket{01d}.
\end{align}
Moreover, with this choice the cyclicity reduces to the condition:
\begin{eqnarray}
    \delta p_i = \delta p_j \equiv \delta p, 
\end{eqnarray}
for all $i$ and $j$, where $\delta p_i = p_{u_i} - p_{d_i}$. 

We arrive with the heat currents
\begin{eqnarray}
    Q_h = d \omega_h \delta p, \quad Q_c = - \omega_c \delta p,
\end{eqnarray}
which leads to the generalized efficiency and work production formulas for $d$-dimensional case:
\begin{eqnarray} \label{d_otto_efficiency}
    \eta_d = 1 - \frac{1}{d} \frac{\omega_c}{\omega_h},
\end{eqnarray}
\begin{eqnarray} \label{d_otto_work}
    W_d = \frac{(d \omega_h - \omega_c)(e^{-d \beta_h \omega_h} - e^{-\beta_c \omega_c})}{(1+e^{-\beta_h \omega_h})(1+e^{-\beta_c \omega_c}) f_d}.
\end{eqnarray}
Above,
\begin{multline}
f_d = \frac{(e^{-d \beta_h \omega_h} - 1) (e^{- \beta_c \omega_c} -1)}{(1-e^{- \beta_h \omega_h})^2}  \\ +  \frac{d (e^{-d \beta_h \omega_h} - e^{- \beta_c \omega_c}) (e^{- \beta_h \omega_h} -1) }{(1-e^{- \beta_h \omega_h})^2} .
\end{multline}


\begin{figure}[t]
    \centering
    \includegraphics[width=0.90\linewidth]{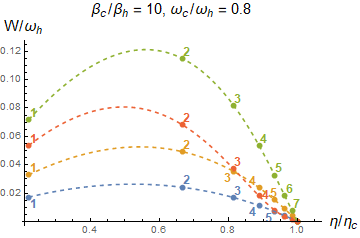}

    \vspace{0.2cm}
    
    \includegraphics[width=0.90\linewidth]{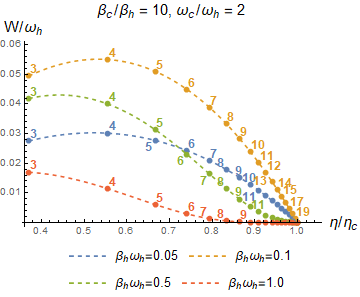}
    \caption{Efficiency vs. work production trade-off. For a fixed ratio of temperatures $\beta_c/\beta_h$, ratio of frequencies $\omega_c/\omega_h$ and hot temperature $\beta_h \omega_h$, the trade-off curve is constructed via changing the dimension of the catalyst $d$ (points on the curve). According to formula \eqref{d_otto_efficiency}, the efficiency increases with growing dimension $d$; however, the growth in work production is also observed from non-catalytic ($d=1$) to catalytic engine $(d=2)$ in the top panel. The range of possible dimensions is given by Eq. \eqref{range}.}
    \label{fig:trade_off1}
\end{figure}

\textit{Regime of operation.}
According to Eq. \eqref{d_otto_efficiency}, the catalyst provides an enhancement in engine efficiency $\eta_d$ provided the machine works in the engine's mode, i.e., whenever the provided work \eqref{d_otto_work} is positive $W_d>0$. The condition boils down to the following inequality: 
\begin{eqnarray}\label{regime_ineq}
    \frac{\beta_h}{\beta_c} < \frac{\omega_c}{d \omega_h} < 1,
\end{eqnarray}
that proves that the engine's efficiency is always smaller than Carnot, i.e., $\eta_d < \eta_c = 1 - \beta_h/\beta_c$.
Equivalently, for fixed bath temperatures we provide the range of all possible dimensions of the catalyst:  
\begin{eqnarray} \label{range}
    d \in \left(\frac{\omega_c}{\omega_h}, \frac{\beta_c \omega_c}{\beta_h \omega_h} \right). 
\end{eqnarray}
(A larger range of parameters for which the engine with $d$-dimensional catalyst is operating was obtained via a more general protocol given in \cite{BiswasNew}).
Fig. \ref{fig:ranges} presents the range of operation  of the engine assisted by the $d$-dimensional catalyst, given by the ratio of frequencies $\omega_c/\omega_h$ and temperatures $\beta_c/\beta_h$, with extensions due to catalyst visible. 
Nevertheless, for every $d>1$, the range for $d$ does not cover the entire range for $d-1$, so that there exist a regime where only standard two-stroke Otto engine works (i.e., with $d=1$), and catalytic approach is ineffective in improving its efficiency or work production.

\begin{figure}[t]
    \centering
    \includegraphics[width=0.90\linewidth]{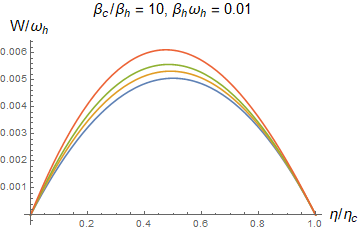}

    \vspace{0.2cm}
    
    
    
    \includegraphics[width=0.90\linewidth]{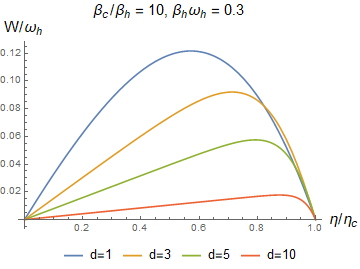}
    \caption{Efficiency vs. work production trade-off. For a fixed ratio of temperatures $\beta_c/\beta_h$, hot temperature $\beta_h \omega_h$ and dimension of the catalyst $d$, the trade-off curve is constructed via changing the ratio of frequencies  $\omega_c/\omega_h$. In the top figure for high temperature $\beta_h \omega_h = 0.01$, the trade-off is favorable for catalytic engines, whereas for lower temperature $\beta_h \omega_h = 0.3$ (bottom panel), a transition to the opposite regime is observed.}
    \label{fig:trade_off2}
\end{figure}

\textit{Efficiency vs. work production.} Finally, one may ask what is the trade-off between efficiency and work production (per cycle) for engines assisted by the catalyst, and how the Carnot limit is approached. In Fig. \ref{fig:trade_off1} and \ref{fig:trade_off2}, we provide an analysis in two different ranges of parameters. In the former, the trade-off is explored via changing the dimension of the catalyst $d$, whereas in the latter via changing the ratio of frequencies $\omega_c/\omega_h$. In Fig. \ref{fig:trade_off1} we observe an increase in work production when going from the non-catalytic ($d=1$) to the catalytic engine (e.g., $d=2$), before the work production goes to zero at Carnot's efficiency point. On the other hand, Fig. \ref{fig:trade_off2} shows that, depending on the temperature regime, maximizion of work production may single out settings without a catalyst. Nevertheless, in the limit of high efficiencies, the catalytic approach can still prevail. 

\emph{Discussion.} 
The letter outlines a microscopic heat engine composed of a pair of two-level systems that completes the thermodynamic cycle in just two strokes. This so-called two-stroke Otto engine operates at the optimal efficiency given by the Otto's formula \cite{BiswasNew}. Our main contribution lies in formulating protocols that exceeds that efficiency by incorporating a catalyst into its construction. 
Notably, we have demonstrated that even a catalyst as basic as a two-level system is sufficient to surpass the Otto efficiency. We have extended this protocol to include a $d$-level system as a catalyst, revealing that the efficiency can be elevated to the $d$-Otto efficiency \eqref{d_otto_efficiency}. Furthermore, we have analytically determined the amount of work generated by such an engine and analyzed the trade-off curve between work and efficiency. The work-efficiency trade-off curve demonstrates that a catalyst enables us to simultaneously exceed the optimal work production and efficiency, which is a particularly intriguing result. 

Our findings pave the way for future research. Using the formalism described in this letter, it is relatively straightforward to enhance the performance of other thermal machines, such as coolers or heat pumps. 
Addressing these questions would not only facilitate the development of more efficient microscopic thermal machines but also provide insights into the deseign of energy-efficient quantum algorithms for computation and information processing tasks (see e.g. \cite{NYHSG}). Another avenue of research emerges from the question: Can construction principles of a two-stroke thermal machine be translated to an autonomous thermal machine that produces work and exchanges heat continuously and simultaneously \cite{Kosloff_Uzdin,MitchisonContemp,discvscont}? Establishing such a correspondence would enable the replication of the efficiency of two-stroke thermal machines in a control-free manner, which is more experimentally feasible. For instance, experiments could be designed to implement two-stroke thermal machines using superconducting qubits or within the framework of cavity quantum electrodynamics \cite{NYHSG,Yu19}. Furthermore, establishing the connection between stroke-based discrete thermal machines and autonomous thermal machines would provide estimates for the time required to complete a thermodynamic cycle.

\emph{Acknowledgement:} TB acknowledges Pharnam Bakhshinezhad for insightful discussions and comments during a visit in quantum information and thermodynamics group at TU, Vienna.  The part of the work done at Los Alamos National Laboratory (LANL) was carried out under the auspices of the US DOE and NNSA under contract No.~DEAC52-06NA25396. TB also acknowledges support by the DOE Office of Science, Office of Advanced Scientific Computing Research, Accelerated Research for Quantum Computing program, Fundamental Algorithmic Research for Quantum Computing (FAR-QC) project. MŁ, PM and TB acknowledge support from the Foundation for Polish Science through IRAP project co-financed by EU within the Smart Growth Operational Programme (contract no.2018/MAB/5). 
MH acknkowledges the support by the Polish National Science Centre grant OPUS-21 (No: 2021/41/B/ST2/03207). MH is also partially supported by the QuantERA II Programme (No 2021/03/Y/ST2/00178, acronym Ex-TRaQT) that has received funding from the European Union’s Horizon 2020.

\bibliography{literature}
\end{document}